\documentstyle[aps,epsfig]{revtex}

\newcommand{\gm}[1]{\mbox{$\; / \!\!\!\! #1$}}
\begin{document}
\vspace*{0.2in}
\begin{center}
{\LARGE Meson-Baryon Form Factors in Chiral Colour Dielectric Model}\\
\vspace*{0.2in}
 S. C. Phatak \\
\vspace*{0.2in}
{\it Institute of Physics 
Bhubaneswar 751 005}
\end{center}

\vspace{0.2cm}
\begin{abstract}
The renormalised form factors for pseudoscalar meson-baryon coupling  are  
computed in chiral colour dielectric model. 
This has been done by rearranging the Lippmann-Schwinger series for the
meson baryon scattering matrix so that it can be expressed as a baryon pole term
with renormalized form factors and baryon masses and the rest of the terms which
arise from the crossed diagrams. Thus we are able to obtain an integral equation
for the renormalized meson-baryon form factors in terms of the bare form factors
as well as an expression for the meson self energy. 
This integral equation is solved and renormalized meson baryon form factors 
and renormalized baryon masses are computed. The parameters of the model
are adjusted to obtain a best fit to the physical baryon masses. The 
calculations show that the renormalized form factors are energy-dependent and
differ from the bare form factors primarily at momentum transfers smaller than
1 GeV. At nucleon mass, the change in the form factors is about 10\% at zero
momentum transfer. The computed form factors are soft with the equivalent 
monopole cut-off mass of about 500 MeV. The renormalized coupling constants
are obtained by comparing the chiral colour dielectric model interaction 
Hamiltonian with the standard form of meson-nucleon interaction Hamiltonian.
The ratio of $\Delta N\pi$ and $NN\pi$
coupling constants is found to be about 2.15. This value is very close to the
experimental value.  
\end{abstract} 

PACS numbers: 12.39.Ki, 12.39.Fe, 13.75.Gx

\vspace{0.2cm}

\section{Introduction}

The meson-baryon form factors are interesting and useful quantities for several
reasons. For one thing, we expect the form factors to have some bearing on the 
underlying structure of the hadrons. For example, in perturbative chiral quark 
models\cite{CBM} the pion-baryon form factors are directly related to the quark 
wavefunctions in baryons. Form factors are essential in the effective models of
meson-baryon interactions since hadrons, after all,
are not point particles. One also needs these form factors for a consistant 
description of nuclear phenomena. The meson exchange nucleon-nucleon potentials
include phenomenological vertex form factors which are presumably related to the
structure of hadrons. Form factors are also needed in computation of
photo- and electro- production of baryon resonances since, in principle, the 
photon can couple to the virtual charged meson cloud in a baryon. Another 
technical reason for introducing the form factors is that the effective 
meson-baryon interaction models are, generally, not renormalizable and the 
form factors provide the needed cut-off functions.

Usually, the form factors employed in the meson-baryon interaction models are
of monopole type ( $\sim \frac{1}{1-k^2/\Lambda^2}$ ) with $k^2$ being the
four momentum carried by the meson. It must be noted that the monopole form is, 
as such, purely phenomenological and is possibly not related to the underlying
structure of the hadrons. The form factors used in earlier 
calculations\cite{Bonn} were `hard' with the cut-off parameters $\Lambda \geq 1 
GeV$. There are, however, recent calculations\cite{Jain} which indicate that
softer form factors ($\Lambda \sim 500 MeV$) are required to fit $\Delta$ 
production on nuclei. Some indication of the value of cut-off parameter 
$\Lambda$ is available from low energy phenomenology. For example, Thomas and
Holinde\cite{TH} argue that the 3\% descripancy between $pp\pi^0$ and $pn\pi^+$
coupling constants\cite{BCR} is essentially due to the four-momentum variation 
of the $\pi NN$ form  factor between $q^2 = m_\pi^2$ and $q^2 = 0$. With
monopole parametrization of the form factor, $\Lambda$ of about 800 MeV is
required to explain this descripancy. Similar arguments have been proposed by
Coon and Scadron\cite{CS} for $\Lambda \sim 800 MeV$. One should however note 
that these analyses are restricted to relatively small values of $|q^2|$ 
( $\sim (140 MeV)^2$ ) and therefore cannot determine the form factors at
larger momentum transfers. Recently, Saito and Afnan\cite{SA} have investigated
the dependence of triton binding energy on the $\pi NN$ form factor. They 
They compute the three-body $\pi - \pi$ force contribution to the triton 
binding energy. They claim that one can determine the $\pi NN$ form factor
their model and it turns out that their renormalized form factor is softer
with $\Lambda \sim 400 MeV$. On the other hand, Schultz and Holinde\cite{SH} 
fit $\pi N$ scattering data and obtain a form factor having $\Lambda \sim 800 
MeV$. Some attempts have been made to determine the pion-baryon form factor
from QCD. For example, Liu {\it et al.}\cite{Liu} have used quenched lattice 
QCD calculations and extracted $\Lambda$ of 750 MeV. T. Meissner\cite{Mei}, on
the other hand, has used QCD sum rules and obtained $\Lambda$ of about 800 MeV.

In the present work, the renormalised pseudoscalar meson baryon  form factors 
are computed by using the
chiral colour dielectric (CCD) model of baryons\cite{CCD}. The CCD model is 
based on the calculation of Nielsen and Patkos\cite{NP} who showed that `coarse
graining' of QCD the Lagrangian on lattice gives rise to an effective scalar,
colour-neutral field called colour dielectric field. In the CCD model, the
the interaction of the colour dielectric field with the quark and gluon fields
is such that it gives rise to the confinement of these 
objects. The chiral symmetry is restored in the CCD model by introducing the 
interaction of the pseudoscalar meson octet with the quarks\cite{CCD}. 
The CCD model has been used to compute properties of light baryon\cite{CCD} as 
well as $\pi N$ scattering phase shifts\cite{PLL}. Generally the agreement 
between the CCD model results and the experimental data is good. In particular,
a good agreement with the $\pi N$ scattering phase shifts is obtained in the CCD
model. In the $\pi N$ work, the authors showed that solving of the 
relativistic Lippmann-Schwinger equation for pion-nucleon scattering leads to 
the renormalization of baryon masses and  the pion-baryon veritces are
also renormalized due to multiple scattering. However the renormalized vertex 
functions ( or form factors ) were not computed in that work. The present work 
is essentially an extension of the earlier work\cite{PLL} but here we have
concentrated on renormalized form factors and have not compute $\pi N$ 
scattering matrix. The basic idea here is as follows. The CCD model ( or, for 
that matter any other quark-based model with mesons coupling to quarks ) give 
bare meson-baryon form factors and bare baryon masses. These are then dressed 
by meson-baryon interactions. As we shall show later in the 
paper, computation of meson-baryon scattering ( including the pole positions
of stable particles ) leads to the determination of dressed baryon masses 
as well as dressed meson-baryon form factors. This is essentially achieved by
rearranging the terms in the Lippmann-Schwinger series for the meson-baryon
scattering matrix so that one obtains equations for the dressed baryon 
masses ( positions of the poles in the scattering matrix ) and meson-baryon
form factors ( see the following sections for details ). Thus, one can compute
the dressed pion baryon form factors. One should note here that these
dressed ( or renormalized ) form form factors, and not the bare form factors
should be used in nuclear physics calculations. Some saliant features,
of the results of our calculation are
\begin{enumerate}
\item The renormalization of the form factors depends on the energy of 
the meson-baryon system. At zero momentum transfer, the ratio of renormalized
and bare  form factors
varies from about 1.1 to about 1.3 as the energy
is increased from nucleon mass to $\Delta$ mass. 
\item The renormalized form factors differ from the bare ones for meson momenta
of 1 GeV or smaller. There is practically no change at larger meson momenta.
\item The form factors computed in the CCD model are soft. If one wants to fit
the CCD model form factors by a monopole form in the region of 300-700 Mev, 
corresponding $\Lambda$ is about 500 MeV.
\end{enumerate}

At this stage we would like to emphasize that our prescription of computing 
the renormalized form factors should be applicable to models having basic 
baryon-meson interaction as an input. That is, given a model for baryon-meson
interaction Lagrangian with bare baryon masses and form factors, one should 
compute the renormalized form
factors by fitting the data ( cross sections, phase shifts, positions of 
the poles of the scattering matrix etc). These renormalized form factors should 
then be used in other nuclear physics calculations
such as NN potentials, photo- and electro- production of mesons etc. 
In other words, the parameters of the basic baryon-meson model are necessarily 
the bare parameters and should not be obtained
directly from the experiments. If one does this then one is essentially
choosing dressed parameters in the Lagrangian and then one should
use the model at tree level and should not compute meson loops for meson-nucleon
scattering etc since that would amount to double counting. 

The paper is organised as follows. In Section 2, a brief description of  the 
CCD model is given and the bare masses and pion-baryon form factors are 
computed. In Section 3, begining with the Lippmann-Schwinger series, the 
the integral equations for the meson-baryon form factors are obtained.
In Section 4, the results of the calculation are presented and discussed. 

\section{Chiral Colour Dielectric Model}
The CCD model is described and the bare masses of baryons and pseudoscalar 
meson-baryon vertices are computed in this section. The description of the CCD
model is somewhat brief because in this work we want to focus on  
the computation of meson-baryon form factors.
For the details of the CCD model the reader is refered to \cite{CCD,PLL}.

\subsection{Lagrangian} 

The Lagrangian density of CCD model is given by\cite{CCD} 
\begin{eqnarray}
{\cal L}(x)&=& \bar\psi(x)\Big \{ i \gm{ \partial } -
(m_{0}+\frac{m}{\chi(x)} U_{5}) + \frac{g}{2} 
\lambda^c_a \gm{ A }^{a}(x)\Big \}\psi 
+\frac{f^2}{4} Tr ( \partial_{\mu}U
\partial^{\mu}U^{\dagger} ) - \nonumber \\
& & -\frac{1}{2}m^{2}_{\phi} \phi^{2}(x) -\frac{1}{4}
\chi^{4}(x)(F^{a}_{\mu\nu}(x))^{2} 
 + \frac{1}{2}
\sigma^{2}_{v}(\partial_{\mu}\chi(x))^{2}- U(\chi) 
\end{eqnarray}
where $U = e^{i\lambda^f_a\phi^{a}/f}$ and $U_{5} =
e^{i\lambda^f_a\phi^{a}\gamma_{5}/f}$, $\psi(x)$,
$A_{\mu}(x)$, and $\chi(x)$ and $\phi(x)$ are quark, gluon,
scalar ( colour dielectric )and meson fields respectively.
$m$ and $m_{\phi}$
are the masses of quark and meson, $f$ is the pion
decay constant, $F_{\mu\nu}(x)$ is the usual colour
electromagnetic field tensor, g is the colour coupling constant
and $\lambda^c_a$ and $\lambda^f_a$ are the usual Gell-Mann matrices
acting in colour and flavour space respactively.  The flavour
symmetry breaking is incorporated in the Lagrangian through the
quark mass term $(m_{0}+\frac{m}{\chi} U_{5})$, where $m_0= 0$
for u and d quarks. So masses of u, d and s quarks are $m$, $m$
and $m_{0}+ m$ respectively. The meson matrix consists of a
singlet $\eta$, triplet of $\pi$ and quadruplet of $K$.

The self interaction
$U(\chi)$ of the scalar field is assumed to be of the form
\begin{eqnarray}
U(\chi)~=~\alpha B
\chi^2(x)[
1-2(1-2/\alpha)\chi(x)+(1-3/\alpha)\chi^2(x)]
\end{eqnarray}
so that $U(\chi)$ has an absolute minimum at $\chi = 0$ and a secondary
inimum at $\chi = 1$.

We would like to make the following observations on the
Lagrangian of eq(1).
\begin{enumerate}
\item The CCD model Lagrangian is invaraint under chiral
transformations if the mass terms of the pseudoscalar mesons are
dropped.
\item The coupling of the scalar field ($\chi$) to the quark
and gluon fields is such that the quark mass becomes infinite
and the dielectric function ( $\chi^4$ ) vanishes when $\chi$
becomes zero. This means that the quarks and gluons cannot exist
in a region where  $\chi$ becomes zero. That is the confinement of quarks and
gluons is included in the model.
\item The two minima of the self-interaction of the scalar field
(eq(2)) can be identified with the perturbative and physical
vacua of the MIT bag model\cite{bag}. Thus, the CCD model is a dynamical
model generating the bag.
\end{enumerate}

We follow the cloudy bag model approach\cite{CBM} in the present work and 
treat the gluon and meson interactions with the quarks perturbatively. 
Therefore we expand the Lagrangian in powers of $1/f$ and keep the terms 
up to order $1/f$ in the Lagrangian of eq(1).
With this approximation the CCD model Lagrangian becomes,
\begin{eqnarray}
{\cal L}(x)&=& \bar\psi(x)\Big \{ i\gm{\partial}-
(m_{0}+\frac{m}{\chi(x)}(1 + \frac{i}{f}\lambda_a\phi^a(x) +\frac{g}{2}
\lambda_{a}\gm{A}^{a}(x)\Big \}\psi \nonumber \\
& & + \frac{1}{2} (\partial_\mu \phi_a(x))^2 -
\frac{1}{2}m^{2}_{\phi} \phi^{2}(x) -\frac{1}{4}
\chi^{4}(x)(F^{a}_{\mu\nu}(x))^{2}
+ \frac{1}{2}
\sigma^{2}_{v}(\partial_{\mu}\chi(x))^{2}- U(\chi)
\end{eqnarray}

The parameters of the CCD model are quark masses ( $m$ and $m_0$ ), the `bag
constant' ( $B$ ), strong coupling constant ( $\alpha_s = g^2/4\pi$ ), pion
decay constant ( $f$ ), glueball ( or dielectric field ) mass ( $m_{GB}$ ) 
and the constant $\alpha$ in $U(\chi)$. Of these parameters, value of $\alpha$ 
is chosen to be 24 since from our earlier calculations\cite{CCD}
the results are not sensitive to it. To bigin with, we choose the experimental 
value of the pion decay constant ( $f = 93 MeV $ ) in our calculations. As we 
shall see later, the value of $f$ required to fit pion-nucleon coupling constant
is very close to $93 MeV$.
Thus we are left with five free parameters to be adjusted. In our earlier 
calculations\cite{CCD} we found that a reasonably good fit to the baryon masses
is obtained for $m$ and $B^{1/4}$ ranging between 100 and 140 MeV. We therefore 
choose $m$ and $B^{1/4}$ in this range and adjust $m_{GB}$, $\alpha_s$ and $m_0$
to fit nucleon, $\Delta$ and $\Lambda$ masses. Computed masses of other octet 
and decuplet baryons are within few percent of their experimental masses.

\subsection{Bare Baryon States}

The equations of motion of quark, gluon and  colour
dielectric fields can be obtained from the Lagrangian
of eq(3). Since we shall be treating the gluon interactins perturbatively,
the quark and colour dielectric field equations of motion are solved
self-consistantly. The gluon as well as pseudoscalar meson interactions are
then treated perturbatively. This approach is similar to the one followed in
the cloudy bag model calculations\cite{CBM}. Thus the equations of motion for 
the quark and dielectric field are,

\begin{eqnarray}
(\epsilon_i - m_0 -\frac{m}{\chi(r)} ) g_i(r) & = & - f'_i(r) -
\frac{2}{r} f_i(r) \nonumber \\
(\epsilon_i - m_0 - \frac{m}{\chi(r)} ) f_i(r) & = & g'_i(r)
\end{eqnarray}
and
\begin{eqnarray}
\chi''(r) + \frac{2}{r}\chi'(r) -\frac{2B\alpha}{\sigma^2_v}
\chi(r) \Big [ 1 - 3 ( 1 - 2/\alpha) \chi(r) + 2 ( 1 - 3/\alpha
) \chi^2(r) \Big] 
+\sum_i \frac{N_i m_i}{\sigma^2_v \chi^2(r)}(
g_i^2(r)-f_i^2(r)) = 0.
\end{eqnarray}
We have assumed spherical symmetry in obtaining these equations. 
The equations of motion of quark and dielectric fields are solved 
self-consistantly with the boundary conditions that as $ r \rightarrow \infty$, 
$\chi(r), g_i(r), f_i(r) \rightarrow 0$, $\chi(r)$ and $g_i(r) \rightarrow $ 
constant and $f_i(r) \rightarrow 0$ as $r \rightarrow 0$.
The bare baryon states are constructed by putting three quarks of appropriate
flavour in $1s_{1/2}$ orbital. Thus the baryon wavefunction is a product of
symmetric space wavefunction, symmetric spin-flavour wavefunction and 
antisymmetric colour wavefunction\cite{bag}. The mass of this state 
is computed by
evaluating the matrix element of the Hamiltonian
\begin{equation}
H_0 =\int d^3x \Big [ \sum_i \Psi^\dagger \Big ( -i\vec \alpha \cdot \nabla
+ \frac{m}{\chi}+m_0 \Big ) \Psi + \frac{\sigma^2_v}{2}( (\nabla \chi)^2 +
\Pi^2 ) + U(\chi) \Big ] 
\end{equation}
in the bare baryon state. Here $\Psi$ is an annihilation operator of a quark in 
the state computed in eq(4) and $\Pi$ is the momentum conjugate to the 
dielectric field $\chi$. Method of coherent states\cite{Willets} has been used 
to better account for the energy associated with the dielectric field. To this 
the colour-magnetic interaction between the 
quarks is added perturbatively ( see ref\cite{CCD,Willets1} for details). 
The bare mass of the baryon is then
\begin{eqnarray}
M_B^0 = <B(\vec 0)|H_0|B(\vec 0)> + E_{M}
\end{eqnarray}
where $|B(\vec 0)>$ is the bare baryon state having zero momentum. This state is
constructed by using Peierls-Yoccoz projection technique\cite{PY,Willets2}.
It is convinient to express $M_B^0$ as
\begin{eqnarray}
M_B^0 = m_{GB} ( C_1^B + \alpha_s C_2^B )
\end{eqnarray}
where $C_1^B$ depends on the number of strange quarks in the baryon and $C_2^B$
depends on the spin-flavour wave function of the baryon. ( $C_1^B$ and $C_2^B$
are, of course functions of other parameters of the model. The mass $M_B^0$ 
defined above is the bare mass of baryon B since it does not include the meson
self energy corrections. In order to obtain the physical mass of the baryon one 
should determine the position of the pole of the T-matrix in the appropriate 
meson-baryon channel\cite{PLL}. Clearly, this mass is different from $M_B^0$.
In our calculations we treat $m_{GB}$, $\alpha_s$ and $m_0/m_{GB}$ as free 
parameters and obtain a best fit to the baryon masses. 

\subsection{Bare Meson Baryon Vertex}

The meson baryon interaction Hamiltonian obtained from the Lagrangian of eq(3)
is

\begin{eqnarray}
H_{int} = \frac{i}{f} \int d^3x \frac {m}{\chi(x)} \overline \Psi (x) \gamma_5 
\lambda_i \Psi(x) \phi_i(x)
\end{eqnarray}
In order to compute the bare meson-baryon form factors we quantize the meson 
fields and evaluate the matrix element of $H_{int}$ between bare baryon states
defined in the previous section.  With this, the interaction Hamiltonian can
be written as
\begin{eqnarray}
H_{int} &=& \frac {i}{f} \sum_{B,B'} |B'><B| 
   \int \frac{d^3kd^3x}{\sqrt{16 \pi^3 \omega_\phi(k)}} 
e^{i\vec k \cdot \vec x}  \Big [ <B'| 
\overline \Psi (x) \gamma_5 \lambda_i \Psi(x) \frac{m}{ \chi(x)}  a_i(\vec k) 
|B> + h. c. \Big ] 
\end{eqnarray}

It is convinient to use angular momentum formalism for evaluating the
spin-flavour matrix elements. The details of the calculation are described in
Appendix A. The end result of the calculation is
\begin{eqnarray}
H_{int} &=& \frac {i}{f} \sum_{B,B'} |B'><B|\frac{\alpha_{BB'\phi}}
{\sqrt{16 \pi^3}}
<T_B, T_\phi, t_B, \mu|T_{B'},t_{B'}><S_B,1,s_B,\nu|S_{B'},s_{B'}> \nonumber \\
 & & (-1)^{\nu + i_\phi} \int \frac{d^3k}
{\sqrt{\omega_\phi(k)}}( a_{-\mu}(\vec k)-a^\dagger_\mu
(\vec k)
) u^0_{BB'\phi}(k)  k_{1,-\nu}
\end{eqnarray}
where the quantities in angular brackets $<\cdots>$ are the usual SU(2) 
Clebsch-gordon coefficients, $S_B$ and $T_B$ are spin and isospin of barion B,
$s_B$ and $t_B$ are their projections, $\mu$ is the isospin projection of meson
$\phi$ and $\alpha_{BB'\phi}$ are the reduced matrix
elements. The phase factor $i_\phi$ is defined in the Appendix A.
$\alpha_{BB'\phi}$ are defined in Table 1. The bare form factor 
$u^0_{BB'\phi}(k)$ is evaluated in the `brick-wall' frame so that the
baryons $|B>$ and $|B'>$ have momenta $|\vec k /2>$ and $|- \vec k /2>$
respectively. The momentum states of the baryons are constructed using 
Peierls-Yoccoz projection technique\cite{PY,Willets1}.
\pagebreak
\begin{table}
\caption{The reduced matrix elements $\alpha_{BB'\phi}$ are given in this table.
The matrix elements for K and $\eta$ mesons are given in curly brackets and 
square brackets respectively.}
\begin{tabular}{|c|cccccccc|}
 & N & $\Lambda$ & $\Sigma$ & $\Xi$ & $\Delta$ & $\Sigma ^*$ & $\Xi ^*$ & 
$\Omega$ \\
\hline
& & & & & &  & & \\
N & 5 [1] & \{$-\sqrt{18}$\} & \{$-\sqrt{2\over3}$\} & & $\sqrt{8}$ & 
\{$-\sqrt{8\over3}$\} & & \\
& & & & & &  & & \\
$\Lambda$ & \{ 3 \} & [-2] & -2 & \{1\} & & 2 & \{ -2 \} & \\
& & & & & &  & & \\
$\Sigma$ & \{ -1 \} & $\sqrt{12}$ & $\sqrt{32\over3}$ [2] & \{ -5 \} & \{ 
$-\sqrt{8}$ \} & $\sqrt{8\over3}$ [2] &  \{ -2 \} & \\
& & & & & &  & & \\
$\Xi$ & & \{ $\sqrt{2}$ \} & \{ $\sqrt{50\over3}$ \} & -1 [-3] & & \{ -$\sqrt{
8\over3}$ & 2 [ 2 ] & \{ -4 \} \\
& & & & & &  & & \\
$\Delta$ & $\sqrt{32}$ & & \{ $-\sqrt{64\over3}$ \} & & 5 [ $\sqrt{5}$ ] & \{
$-\sqrt{40\over3}$ \} & & \\
& & & & & &  & & \\
$\Sigma ^*$ & \{$\sqrt{8}$\} &$\sqrt{24}$ & \{ $\sqrt{16\over3}$ [ -$\sqrt{8}$ ]
& \{ $-\sqrt{8}$ & \{ $\sqrt{10}$ & $ \sqrt{40\over3} $ & \{ $\sqrt{20}$ & \\ 
& & & & & &  & & \\
$\Xi ^*$ & & \{ 4 \} & \{ $-\sqrt{16\over3}$\} & \{$-\sqrt{8}$\} [ -$\sqrt{8}$ ]
& &\{$\sqrt{40\over3}$ \}&$\sqrt{40\over3}$[ $-\sqrt{5}$ ] &\{ $\sqrt{20}$ \} \\
& & & & & &  & & \\
$\Omega$ & & & & \{ -4 \} & & & \{ $\sqrt{10}$ \} & [ $-\sqrt{20}$ ] \\
& & & & & &  & & \\
\end{tabular}
\end{table}

\section{Meson-Baryon Scattering}

Given the meson-baryon interaction Hamiltonian of the previous section, one can
compute the tree-level meson-baryon scattering matrix. It involves pole and 
crossed terms shown in Fig. 1. These two terms constitute the driving term or
potential for meson-baryon scattering. To obtain a unitary scattering matrix and
to include multiple scattering effects the relativistic Lippmann-Schwinger
equation 

\begin{eqnarray}
T_{\beta \alpha}(\vec k',\vec k)
        & = & V_{\beta \alpha}(\vec k',\vec k) 
 + \sum_\gamma\int d^3p
        \frac{ V_{\beta \gamma}(\vec k',\vec p)T_{\gamma \alpha}(\vec p,\vec k)}
        {E + i\epsilon -E_\gamma(p)}. \label{T}
\end{eqnarray}
is solved. In this equation the meson-baryon channel ( including the spins and 
flavours of the particles ) is represented by the subscripts and $E_\gamma(p) =
\sqrt{p^2+m^2_{B_\gamma}}+\omega_{\phi_\gamma}(p)$. The potential $V$ above is 
computed from the diagrams of Fig. 1. 
( see Phatak, Lu and Landau\cite{PLL} for details ). 

\begin{figure}[h]
\epsfxsize=14.5cm
\centerline{\epsfbox{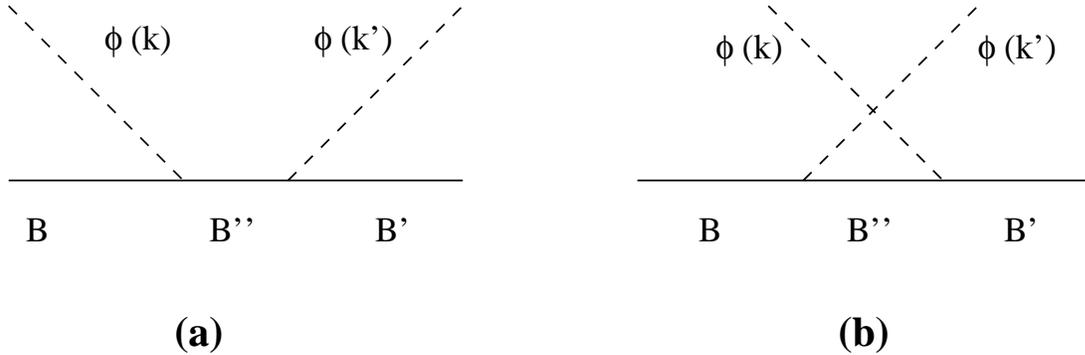}}
\caption{s channel (a) and u channel (b) pole terms ( direct and crossed terms
respectively ) contributing to the meson-baryon
potential}
\end{figure}

We would like to make several points regarding the T-matrix equation given 
above (eq(12)).
\begin{enumerate}
\item The T-matrix given above describes the scattering of one of the octet
of the pseudoscalar mesons from the octet or decuplet of baryons. Thus it
represents resonance production, production of strange mesons and baryons etc.
\item The channel energies $E_\gamma$ in the Lippmann-Schwinger equation
above are defined in terms of the physical baryon masses since these are the 
physically observable states. However the masses of baryons in the
intermediate state of the pole and crossed terms are the bare masses defined
in the previous section.
\item The poles of the T-matrix in appropriate spin-flavour channel of the
meson-baryon system should coincide with the physical masses of
the baryons. This means that the parameters of the CCD model ( basically 
the glueball mass $m_{GB}$, strong coupling constant $\alpha_s$ and $m_0$ )
are to be adjusted to obtain the best fit to the physical baryon masses defined 
by the pole positions of the T-matrix.
\end{enumerate}

Given the potential $V_{\beta \alpha}(\vec k',\vec k)$, one can solve the 
Lippmann-Schwinger equation and obtain the phase shifts and cross sections for
meson-baryon scattering. For this purpose, it is best to expand the T-matrix
equation in spin-isospin channels of the meson-baryon system\cite{PLL} and solve
partial wave Lippmann-Schwinger equation. As mention earlier, when one solves
the Lippmann-Schwinger equation renormalization of baryon masses as well as the 
form factors is already included. However one cannot compute the renormalized
form factors using this procedure. 
Since we are interested in computing the renormalized
form factors in the present work, we do not want to solve the Lippmann-Schwinger
equation as such. Instead, we express the Lippmann-Schwinger equation as a 
multiple scattering series. From this series we obtain integral equations for 
the renormalized form factors by summing up parts of the series. We also obtain 
the expressions for the renormalized masses.
The details of this exercise are outlined in Appendix B. 
The result is an integral equation for the renormalized form factor
$u_{BB'\phi}(k)$
\begin{eqnarray}
u_{BB'\phi}(k) &=& u^0_{BB'\phi}(k) +  \sum_{\phi 'B'',B'''} (-1)^I
\hat S_{B''} \hat S_{B'''} \hat T_{B''} \hat T_{B'''}
 \frac{\alpha_{BB'''\phi '} \alpha_{B'''B''\phi} \alpha_{B''B'\overline \phi '}}
{12 \pi^2 f^2 \alpha_{BB'\phi}} W(T_B T_{B'''} T_{B''} T_{B'};T_\phi T_{\phi '})
\nonumber \\
& &W(S_B S_{B'''} S_{B''} S_{B'};11)
 \int \frac{k'^4dk'
u^0_{BB'''\phi '}(k') u^0_{B'''B''\phi}(k) u_{B'' B' \overline \phi '}(k')}
{\omega_{\phi '}(k')(E-E^0_{B'''}-\omega_{\phi '}(k') - \omega_\phi(k))
(E-E_{B''}-\omega_{\phi '}(k'))}
\end{eqnarray} 
where $\hat a = \sqrt{2 a + 1}$, $E^0_{B}(k)=\sqrt{k^2+(M^0_{B})^2}$, 
$E_{B}(k)=\sqrt{k^2+M_B^2}$, 
$\omega_{\phi}(k)=\sqrt{k^2+m_\phi^2}$, $M_B$ and $M_B^0$ are renormalized 
and bare masses of baryon B respectively, $I=\Delta_\phi + \Delta_{\phi'} - 
S_B - S_{B'} - T_B - T_{B'}$, 
$S_B$ and $T_B$ are the spin and
isospin of baryon $B$ respectively and $W(\cdots)$'s are the usual Rakah
coefficients. The equation for the renormalised propagator of a baryon B is 
\begin{eqnarray}
G(E) & = & \frac{1}{E - M^0_B} + \frac{1}{E - M^0_B} \Sigma (E)G(E) \nonumber \\
     & = & \frac{1}{E - M^0_B - \Sigma (E)}
\end{eqnarray}
where the self energy $\Sigma (E)$ is
\begin{eqnarray}
\Sigma (E) &=&   \sum_{B'\phi}\frac{(-1)^{T_{B'}+S_{B'}-T_B-S_B-\Delta}}
{12 \pi^2 f^2} 
\frac{ \hat S_{B'} \hat T_{B'}}{ \hat S_B \hat T_B}
\alpha_{BB'\phi}\alpha_{B'B\overline \phi} 
  \int \frac{k^4dk u^0_{BB'\phi}(k) u_{B'B\overline \phi}(k)}
{\omega_\phi(k)(M_B - E_{B'} - \omega_\phi(k))}
\end{eqnarray}
where $\Delta$ is 0, -1/2, 1/2, 0 for $\pi$, $K$, $\overline K$ and $\eta$
respectively. The renormalized propagator has a pole at $E =  M^0_B + 
\Sigma (E)$. Thus the renormalized mass of baryon $M_B$ is given by
\begin{eqnarray}
M_B =  M^0_B + \Sigma (M_B). 
\end{eqnarray}
Note that eqs(13-16) are coupled equations since renormalized masses and form
factors appear in these equations. Thus these equations need to be solved 
self-consistantly.
 
We can now use the renormalized form factors and propagators in the expression
for the scattering matrix. We then get,
\begin{eqnarray}
T_{B\phi,B'\phi '}(\vec k, \vec k';E) = T^{P}_{B\phi,B'\phi '}
(\vec k, \vec k';E)+ T^{C}_{B\phi,B'\phi '}(\vec k, \vec k')
\end{eqnarray}
where
\begin{eqnarray}
T^{P}_{B\phi,B'\phi '}(\vec k, \vec k';E)= \sum_{B''}
\alpha_{BB''\overline \phi} \alpha_{B''B'\phi'} \frac{u_{BB''\overline \phi}(k) 
u_{B''B'\phi'}(k')
}{E-M_{B''}}
\end{eqnarray}
and
\begin{eqnarray}
 T^{C}_{B\phi,B'\phi '}(\vec k, \vec k')&=& 
 V^{C}_{B\phi,B'\phi '}(\vec k, \vec k')+ \sum_{B'',\phi ''} \int d^3p \frac
{V^{C}_{B\phi,B''\phi ''}(\vec k, \vec p) 
T^{C}_{B''\phi'',B'\phi '}(\vec p, \vec k')}{E-E_B(p)-\omega_\phi(p)}
\end{eqnarray}
Here $ V^{C}_{B\phi,B'\phi '}(\vec k, \vec k')$ is the crossed term shown in Fig
1. A diagramatic representation of the T-matrix is shown in Fig. 2.  

\begin{figure}[h]
\epsfxsize=14.5cm
\centerline{\epsfbox{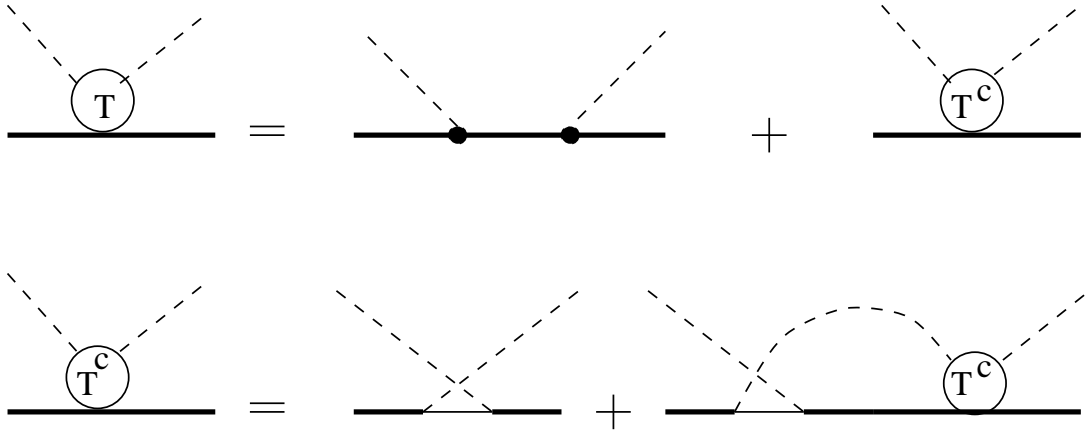}}
\caption{The graphical depiction of T-matrix equation. The first line 
corresponds to eq() of text. The second line depicts the integral equation
for $T^{C}$. The thick (thin) lines represent baryons with physical
( bare ) masses and the broken lines represent mesons. Filled circles represent
renormalized form factor.}
\end{figure}

The interpretation of eq(17) is obvious. The first term represents the pole
or resonance contribution which is given in terms of {\bf physical} masses of
resonances and {\bf renormalized} form factors or vertex functions. The rest of 
the contribution is lumped into the second term which can be considered as a
background. Clearly, if our model is extended to include higher order terms
or exchanges of heavier mesons, these will predominantly affect the background 
term.

Usually the T-matrix is computed by solving the Lippmann-Schwinger equation 
using a suitable numerical technique ( eg matrix inversion method\cite{HL} )
or by summing the Lippmann Schwinger series numerically to a desired accuracy.
Alternatively, one can now compute the renormalized form factors and propagators
and use them to evaluate the T-matrix. The computation of the pole term of the
scattering matrix is very simple in this case. For the crossed term 
$T^{C}_{B\phi,B'\phi '}(\vec k, \vec k')$ however one has to solve an 
integral equation (eq(19)). Our computations show that the crossed term of the 
T-matrix can be computed iteratively and the convergence is rapid. We also
find that the dominant contribution to the scattering matrix comes from the pole
term. One should note that the computation of the renormalized form factor
already includes a large number of terms involving the crossed diagram of the 
meson-baryon potential. It is therefore not
surprising that $T^{C}_{B\phi,B'\phi '}(\vec k, \vec k')$ appearing in
the T-matrix equation (eq(19)) is somewhat less important, particularly near the
resonance.

Note that pole term of the T-matrix $T^{P}_{B\phi,B'\phi '}(\vec k, \vec k';E)$
defined in eq(18) has a singularity 
at $E=M_{B''}$. If this singularity occures at physical energy, it means that
the particle $B''$ can decay and that should be reflected in the mass of the
particle. That is the physical mass defined above should have an 
imaginary part corresponding to the width of the baryon $B''$. 

The renormalised form factors are computed by solving eq(13) above iteratively.
We find that the iterative procedure does converge and the convergence is 
reached within
fifteen iterations. Note that the integrand of the form factor integral equation
has singularities due to the energy denominators and the integral should be 
regularised properly. We have used principal value prescription for this 
purpose. The results of the calculation are discussed in the following Section.

\section{Results and Discussions}

In this Section we shall discuss the results of our calculation. The precedure 
adopted for the calculation is as outlined below. The coupled quark and 
dielectric field equations are first solved and bare baryon states 
are constructed. The bare baryon masses are computed by evaluating the matrix
element of the Hamiltonian (eq(6)) in the momentum-projected bare baryon states
and including the colour magnetic interaction. The baryon masses are essentially
functions of $m_{GB}$, $\alpha_s$ and $m_0$ ( $m$ and $B^{1/4}$ are kept fixed
for a given parameter set ). Also, using the momentum-projected baryon 
baryon wavefunctions the bare form factors ( vertex functions ) for baryon-meson
coupling are computed. These are then used in eq(13) and the renormalised form
factors are computed. In principle, one could use matrix inversion 
method\cite{HL} for this purpose. However we find that the dimensionality of
the matrices is prohibitively large because we are considering coupling to all
the octet of pseudoscalar mesons. We therefore solve eq(13) iteratively. We find
that the convergence is reached after 10-15 iterations. Using the renormalised
form factors the renormalised masses are computed by using eq(16). The 
parameters of CCD model ( $m_{GB}$, $\alpha_s$ and $m_0$ ) are then adjusted to
fit the renormalised masses to the physical masses. 

For the comparison with the experimental data we consider the $\pi NN$ and 
$\pi N \Delta$ coupling constants at nucleon and delta masses respectively.
Conventionally, in terms of the nonrelativistic $\pi NN$ 
interaction Hamiltonian,
the $\pi NN$ coupling constant $f_{\pi NN}$ is defined as\cite{CBM}
\begin{eqnarray}
H_{int} = \frac{i f_{\pi NN}}{m_\pi\sqrt{2}\pi} \int \frac{d^3k}{\sqrt{2 
\omega_\pi(k)}} v(k)[\sigma \cdot
\vec k  \vec \tau \cdot \vec a(\vec k) + h. c. ]
\end{eqnarray}
where $\sigma$ and $\tau$ are nucleon spin and isospin operators respectively, 
$v(k)$ is the form factor defined to be unity when $k = im_\pi$ and $\vec 
a(\vec k)$ is the annihilation operator for pion field. Experimental value of 
$f_{\pi NN}$ is 0.279 ( $ f^2_{\pi NN} = 0.078$ ). Comparing the interaction 
Hamiltonian of eq(11), we have
$f_{\pi NN} = \frac{\alpha_{NN\pi}u_{NN\pi}(k=0)m_\pi}{6\sqrt{\pi}f}$ where the
renormalized form factor is calculated at $E=m_N$, the nucleon mass. Similarly,
$f_{\pi N\Delta} = \frac{\alpha_{N\Delta \pi}u_{N\Delta \pi}(k=0)m_\pi\sqrt{2}}
{5\sqrt{\pi}f}$.

Computations have been done for a number of parameter sets of CCD model. The 
results are quite similar for all these sets so long as $m_{GB}$ is between
800 and 1200 MeV. Therefore we have chosen one of the representative parameter 
set ( $m_{GB}=978.6$, $\alpha_s =0.438$, $B^{1/4} =122.3$, $m =122.3$ and
$m_0 =210.1$ ) for a detailed discussion. The pion decay constant $f$ has been
fixed at its physical value ( 93 MeV ). The bare and renormalized $\pi NN$ and
$\pi N\Delta$ form factors evaluated at nucleon delta masses are displayed in
Fig 3. It is clear from this figure that the renormalizes form factors differ 
from the bare form factor at three momentum transfers smaller than 1 GeV. An
inspection of eq(13) shows why this happens. As the momentum transfer increases,
the first term in the denominator of integrand also increases because of 
increase in $\omega_\phi(k)$. As a result, the contribution of the integral to 
the form factor decreases with the increase in the momentum transfer.
Even at momentum transfers smaller than 1 GeV, the change in the form factor is 
about 10 \% for $u_{NN\pi}$ and 30\% for $u_{N\Delta \pi}$. Thus one could 
conclude that the effects of renormalization are relatively minor.

\begin{figure}[h]
\epsfxsize=14.5cm
\centerline{\epsfbox{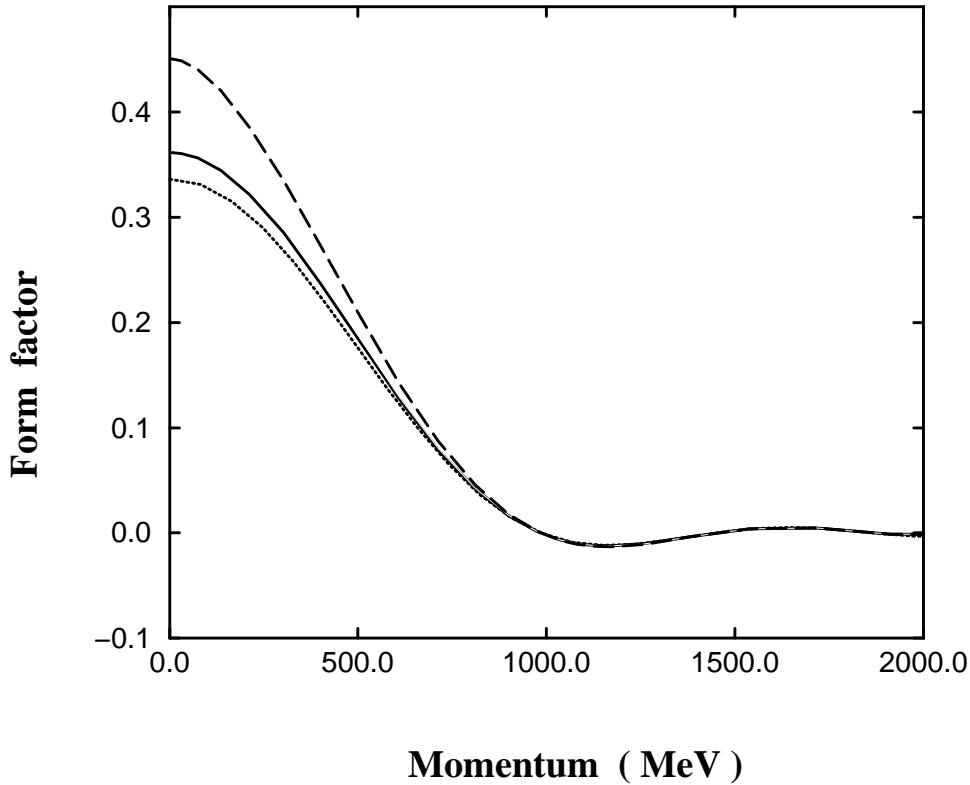}}
\caption{Renormalized and bare form factors {\it vs} momentum in units
of glueball mass. The different curves are 
for $u_{NN\pi}$ ( continuous line ), $u^0_{NN\pi}$ ( dotted line ) and
$u_{N\Delta \pi}$ ( dashed line ). $u^0_{N\Delta \pi}$ is same as $u^0_{NN\pi}$.
}
\end{figure}

We can extract equivalent monopole form factors from the computed renormalized 
form factors. If, for example, $u_{NN\pi}$ is expanded in power series near
$k = 0$, one obtains $\Lambda \sim 800 MeV$. This value of $\Lambda$ is about 
same at the one obtained by various authors\cite{TH,CS,SH,Liu,Mei}. However 
one must note that as function of momentum transfer, the monopole form factor 
falls off more slowly than the CCD model form factor. We find that a better fit
to the CCD model form factor is obtained by using dipole or Gausian form. In 
fact, we find that if we replace the bare CCD model form factor by a dipole or
Gaussian form factor, we get practically the same results.
If, on the other hand, one wants that the monopole form factor should
be close to $u_{NN\pi}(k)$ for $k \sim  300-500 MeV$, $\Lambda \sim 500 MeV$.

The computed value of the $\pi NN$ coupling constant is 0.254. This value is 
about 9\% smaller than the experimental $f_{NN\pi}$. This, by itself, can be 
considered as a success of the CCD model. The $f_{NN\pi}$ as well as the baryon
can be fitted by decreasing the pion decay constant $f$ to 86 MeV ( a 7\% 
decrease ). It may be mentioned  here that this
change in $f$ does not affect the computed properties of baryons very much. 

Let us now consider the ratio of $\pi N\Delta$ and $\pi NN$ coupling
constants evaluated at nucleon and delta masses respectively. This ratio is 
$6\sqrt{2}/5 = 1.697$ for bare coupling constants, as dictated by 
$SU(3)_{flavour}$ symmetry. For renormalised coupling constants this ratio is 
about 2.15, which is very close to the experimental value of 2.14\cite{BCC}.
We believe that this is a very important result emerging from  our calculation. 
This result implies that, to begin with, the quark model value of the  ratio,
which is about 25\% off the experimental value, is itself a good starting point.
In other words, the bare coupling constants in the CCD  model themselves are
in qualitative agreement with the experimental values. The renormalization of 
these quantities, which should be done in any case, is able to produce a very
good agreement with the experimental result. Incidently, we would like to note 
that when $f=86 MeV$
the $\pi NN$ coupling constant agrees with the experimental value and
the ratio of $\pi N\Delta$ and $\pi NN$ coupling constants still remains close
to 2.15. 

\begin{figure}[h]
\epsfxsize=14.5cm
\centerline{\epsfbox{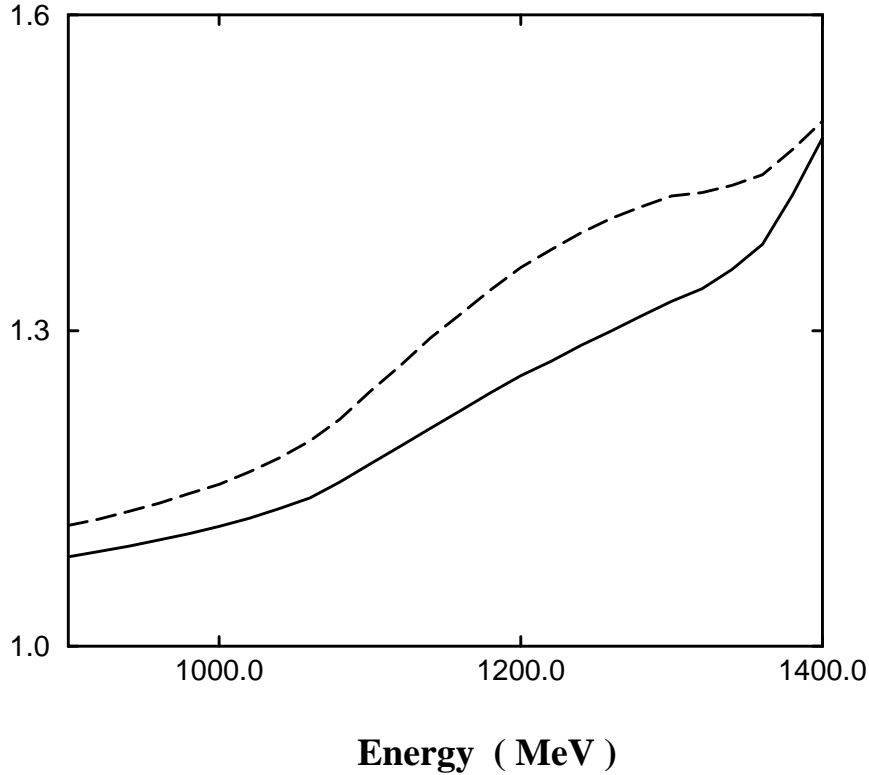}}
\caption{The ratio of renormalized and bare form factor at zero momentum 
transfer as a function of energy. The continuous curve is for $NN\pi$ and 
the dashed curve is for $N\Delta \pi$.
}
\end{figure}

The energy dependence of $f_\pi NN$ and $f_\pi N\Delta$ is shown in 
Fig. 4. The figure shows that the coupling constants increase with the energy 
in the energy range shown.

\section{Conclusions}

The chiral colour dielectric model has been used in this work to compute 
renormalized meson-baryon form factors. This has been done by writing the 
Lippmann-Schwinger equation in a series form and deriving the integral equation
for the renormalized form factor. We find that the renormalization effects
are primarily restricted to the meson momenta smaller than 1 GeV. The increase 
in the form factor at zero meson momentum is about 10\% at nucleon mass and 
about 30\% at $\Delta$ mass. The renormalized $NN \pi$ coupling constant is 
about 10\% smaller than the experimental value if $f$, the pion decay constant
is chosen to be 93 MeV. The ratio of $N\Delta \pi$ and $NN\pi$ coupling 
constants is found to be 2.15, which is close to the experimental value. This 
ratio for bare coupling constants is 1.697. Thus the renormalization effects
yield a good agreement between experiment and theory. The computed form factors
are soft with the equivalent monopole parameter $\Lambda$ of about 800 MeV if 
fitting is done at small momentum transfers. However, the CCD model form factors
decrease faster than the monopole form factor for large momenta and the CCD 
form factors are closer to Gaussian or dipole forms.

Strictly speaking, the computations done in this work are not fully relativistic
although relativistic expressions for energies of the baryons are used. For a 
completely relativistic calculation, one will need to use relativistic 
propagators for baryons and use a relativistic integral equation, such as 
Blankenbeckler-Sugar equation instead of relativistic Scrodinger equation that 
is used in the present work. Indeed it is technically possible to do this, with 
some increase in the complexity of the problem. However we feel that the  
approach followed in the present work is reasonable since we have restricted 
ourselves to kinetic energies small in comparison with the baryon masses.
Extension of the present calculation to higher energies would require the 
above-mentioned relativistic generalizations 
as well as inclusion of heavier baryon resonances.

We would like to thank Prof. R. H. Landau for useful discussions.

\appendix
\section{Bare Form Factors}

The interaction Hamiltonian is
\begin{eqnarray}
H_{int} = \frac{i}{f} \int d^3x \frac {m}{\chi(x)} \overline \Psi (x) \gamma_5
\lambda_i \Psi(x) \phi_i(x)
\end{eqnarray}
where $\lambda_i$ are the SU(3) GellMann matrices acting on the flavour 
coordinate of the quarks. In terms of the pseudoscalar octet fields $\phi_i$,
we define the $\pi$, $K$ and $\eta$ fields as
\begin{eqnarray}
\phi_{\pi^\mp} & =  \phi_{1,\mp 1} & =  \mp {1\over\sqrt{2}} ( \phi_1
\pm i \phi_2) \nonumber \\
\phi_{\pi^0} & =  \phi_{1,0} & =  \phi_3 \nonumber \\
\phi_{K^-} & =  \phi^{(1)}_{1/2,1/2} & =  {1\over\sqrt{2}} ( \phi_4 + i
\phi_5 ) \nonumber \\
\phi_{K^+}  & =  \phi^{(2)}_{1/2,-1/2} & =  {1\over\sqrt{2}} ( \phi_4 - i
\phi_5 ) \\
\phi_{\overline K^0} & =  \phi^{(1)}_{1/2,-1/2} & =  {1\over\sqrt{2}} 
( \phi_6+ i \phi_7 ) \nonumber \\
\phi_{ K^0} & =  \phi^{(2)}_{1/2,1/2} & =  {1\over\sqrt{2}} ( \phi_6 -i 
\phi_7 ) \nonumber \\
\phi_\eta & =  \phi_{0,0} & =  \phi_8. \nonumber 
\end{eqnarray} 
Extra negative sign for $\pi^-$ field is introduced to conform with the usual
definition of vectors in spherical basis and is different from usual definition
of pion fields. 
With these definitions of meson fields,
\begin{eqnarray}
\lambda_a \phi_a & = & (-1)^\mu \tau_\mu \phi_{1,-\mu} + v_{1/2,\mu} \phi^{(1)}
_{1/2,-\mu} + u_{1/2,\mu} \phi^{(2)}_{1/2,-\mu} + \lambda_8 \phi_{0,0}.
\end{eqnarray}
Quantizing the meson fields we have
\begin{eqnarray}
\phi(x) = {1\over\sqrt{8\pi^3}} \int {d^3k \over \sqrt{2 \omega_i(k)}}
\Big [ e^{i\vec k \cdot \vec x} a_i(\vec k) + e^{-i\vec k \cdot \vec x} 
\overline a^\dagger_i(\vec k) \Big ]
\end{eqnarray} 
where $a_i$ is the annihilation operator of meson i and $\overline a^\dagger_i$
is the creation operator of meson conjugate to i 
The commutation relations for the meson field operators are \\
$[a_{1,\mu}(\vec k),a^\dagger _{1,\nu}(\vec k')]=(-1)^\mu \delta_{\mu,\nu} \delta(\vec k -
\vec k')$, \hspace{0.3cm} $[a^{(i)}_{1/2,\mu}(\vec k), a^{(j)\dagger}_{1/2,\nu}(\vec k')] =
\delta_{i,j} \delta_{\mu,\nu} \delta(\vec k -\vec k')$ \\ and
$[a_{0,0}(\vec k), a^\dagger
_{0,0}(\vec k')] = \delta(\vec k -\vec k')$.
Computation of the matrix
element of the interaction Hamiltonian between bare baryon states gives
\begin{eqnarray}
H_{int} = \sum_{B,B',i} |B'><B| {i \over f \sqrt{8\pi^3}} \int {d^3k \over
\sqrt{2 \omega_i(k)}} \Big [ a_i(\vec k) - \overline a^\dagger_i(\vec k) \Big ]
u^0_{BB'i}(k) <B'| \lambda_i \vec \sigma |B> \cdot \vec k
\end{eqnarray}
The details of computing the bare form factor $u^0_{BB'i}(k)$ is given 
elsewhere\cite{CCD}. Computing the spin-flavour matrix element
$ <B'| \lambda_i \vec \sigma |B>$ and expressing the result in angular momentum
formalism, we get
\begin{eqnarray}
H_{int} & = & \sum {i \alpha_{BB'\phi} \over 4f\pi^{3/2}} \int { d^3k\over \sqrt
{\omega_\phi(k)}}\Big [ a_{\lambda ,-\mu}(\vec k) - a^\dagger_{\lambda,\mu}
(\vec k) \Big ] k_{1,-\nu} u^0_{BB'\phi}(k) \nonumber \\
& & (-1)^{\nu+i_\phi} <T_B, \lambda, t_B, \mu|T_{B'},t_{B'}>
<S_B, 1, s_B, \nu|S_{B'}, s_{B'}>
\end{eqnarray}
Here the summation over spin and isospins of baryons and mesons is implied. Also
$\phi$ implies the meson type ( $\pi$, $K$, $\overline K$ and $\eta$ ) and their
total isospins and projections along z-axis are represented by $\lambda$ and 
$\mu$ respectively. The baryon spin ( isospin ) and its projection is 
represented by $S_B$ ( $T_B$ ) and $s_{B}$ ( $t_B$ ) respectively. The reduced 
matrix element $\alpha_{BB'\phi}$ and the phase factor $i_\phi$ are determined  
by exlpicit calculation of one matrix element, as is usually done. The phase 
factor $i_\phi$ is $\mu$, 0, $1/2-\mu$ and 0 for $\pi$, $K$, $\overline K$ and 
$\eta$ respectively.  

\section{Renormalization}

One can derive the equations for renormalized form factors and masses
by writing the T-matrix equation in a Lippmann-Schwinger series and summing 
parts of the series. This procedure, however, is somewhat cumbersome. On the 
other hand, same conclusions can be arrived at by considering the graphical 
representation of the series. We have therefore shown the diagrams corresponding
to the individual terms of the Lippmann-Schwinger series for the T-matrix
of eq() in Fig(5). The figure shows the diagrams upto $1/f^6$. It should be 
clear that one can easily generalize the conclusions drawn from the analysis of 
these diagrams to the full Lippmann-Schwinger series.

\begin{figure}[h]
\epsfxsize=14.5cm
\centerline{\epsfbox{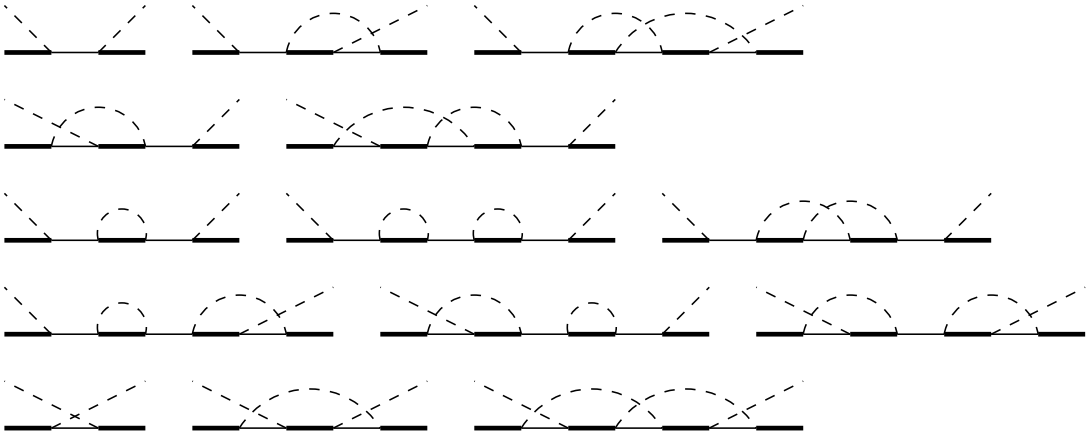}}
\caption{Diagrams ( upto order $1/f^6$ ) contributing to the T-matrix. Thick 
lines represent physical baryons, thin lines represent bare baryons and broken
lines represent mesons} 
\end{figure}

The diagrams in fig(7) are organized as follows. The diagrams in the first line 
renormalize the right-most vertex. The diagrams in the second line renormalize
left-most vertex and the diagrams in the third line renormalize the mass or
the propagator of the intermediate baryon. The diagrams in the fourth line 
are the mixed diagrams representing renormalization of vertices as well as mass.
Finally, the diagrams in the last line are the rest
of the diagrams which do not give rise to renormalization. Generalizing the 
diagrams in the first line to all orders, we obtain an integral equation for
the renormalization of vertex ( form factor ). Diagramatically, this is 
represented in Fig(6) where the filled circle represents the renormalized form
factor.

\begin{figure}[h]
\epsfxsize=14.5cm
\centerline{\epsfbox{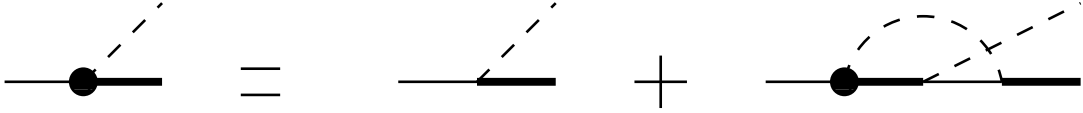}}
\caption{Diagramatic representation of vertex renormalization}
\end{figure}

The corresponding integral equation obtained after some angular momentum algebra
is
\begin{eqnarray}
u_{BB'\phi}(k) &=& u^0_{BB'\phi}(k) +  \sum_{\phi 'B'',B'''} (-1)^I
\hat S_{B''} \hat S_{B'''} \hat T_{B''} \hat T_{B'''}
 \frac{\alpha_{BB'''\phi '} \alpha_{B'''B''\phi} \alpha_{B''B'\overline \phi '}}
{12 \pi^2 f^2 \alpha_{BB'\phi}} W(T_B T_{B'''} T_{B''} T_{B'};T_\phi T_{\phi '})
\nonumber \\
& &W(S_B S_{B'''} S_{B''} S_{B'};11)
 \int \frac{k'^4dk'
u^0_{BB'''\phi '}(k') u^0_{B'''B''\phi}(k) u_{B'' B' \overline \phi '}(k')}
{\omega_{\phi '}(k')(E-E^0_{B'''}-\omega_{\phi '}(k') - \omega_\phi(k))
(E-E_{B''}-\omega_{\phi '}(k'))}
\end{eqnarray}
where $\hat a = \sqrt{2 a + 1}$, $E^0_{B}(k)=\sqrt{k^2+(M^0_{B})^2}$,
$E_{B}(k)=\sqrt{k^2+M_B^2}$,
$\omega_{\phi}(k)=\sqrt{k^2+m_\phi^2}$, $M_B$ and $M_B^0$ are renormalized
and bare masses of baryon B respectively, $I=\Delta_\phi + \Delta_{\phi'} -
S_B - S_{B'} - T_B - T_{B'}$ with $\Delta_\phi$ being 1, 0, 1, 0 for $\pi$, $K$,
$\overline K$ and $\eta$ respectively, $S_B$ and $T_B$ are the spin and
isospin of baryon $B$ respectively and $W(\cdots)$'s are the usual Rakah
coefficients.

\begin{figure}[h]
\epsfxsize=14.5cm
\centerline{\epsfbox{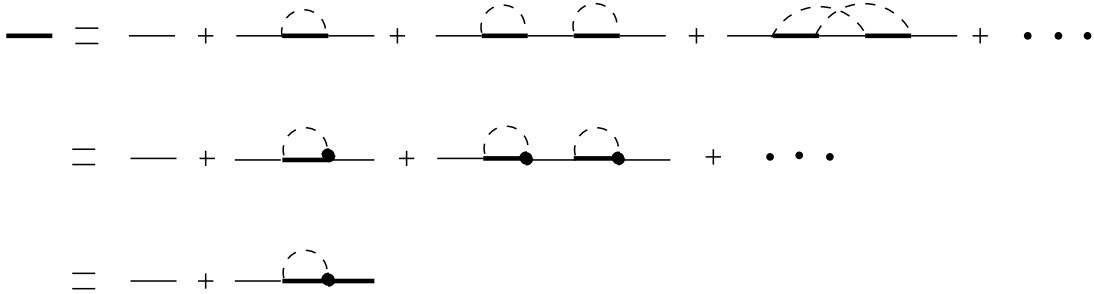}}
\caption{Diagramatic representation of renormalized propagator}
\end{figure}

Now consider the mass renormalization. The relevent diagrams are shown in 
Fig(7).  Clearly, the third line in this diagram is the Schwinger-Dyson 
equation for the propagator with the self energy being given by the meson loop.
The thing to notice, however, is that one of the meson-baryon form factor 
appearing in the self energy is a renormalized one where as the other is a
bare one. Assuming that the computation is done in the frame in which the baryon
is stationary, the renormalized and bare propagators are 
$\frac{1}{E-M^0_{B} - \Sigma(E)}$ and $\frac{1}{E-M^0_{B}}$ respectively.  
The self energy $\Sigma(E)$  is
\begin{eqnarray}
\Sigma (E) &=&  - \sum_{B'\phi}\frac{(-1)^{T_{B'}+S_{B'}-T_B-S_B-\Delta}}
{12 \pi^2 f^2} 
\frac{ \hat S_{B'} \hat T_{B'}}{ \hat S_B \hat T_B}
\alpha_{BB'\phi}\alpha_{B'B\overline \phi} 
  \int \frac{k^4dk u^0_{BB'\phi}(k) u_{B'B\overline \phi}(k)}
{\omega_\phi(k)(M_B - E_{B'} - \omega_\phi(k))}
\end{eqnarray}
where $\Delta$ is 0, -1/2, 1/2, 0 for $\pi$, $K$, $\overline K$ and $\eta$
respectively. 
The mass of the physical baryon $M_{B} = M^0_{B} + \sigma(E = M_B)$.

It should be evident that using the renormalized form factors and propagators
the T-matrix can be written in the form given in eqs(16-18) and Fig(2).

\end{document}